\documentclass[11pt]{article}

\usepackage[version=4]{mhchem}
\usepackage[a4paper, margin=1in]{geometry}

\usepackage[T1]{fontenc}
\usepackage[utf8]{inputenc}
\usepackage{lmodern}
\usepackage{xcolor, soul}
\sethlcolor{yellow!35}
\usepackage{amsmath, amssymb, amsfonts}
\usepackage{siunitx}

\usepackage{graphicx}
\graphicspath{{Figures/}}
\usepackage{caption}
\usepackage{subcaption}

\usepackage{booktabs}

\usepackage{authblk}

\usepackage[numbers,sort&compress]{natbib}
\usepackage[colorlinks=true, linkcolor=blue, citecolor=blue, urlcolor=blue]{hyperref}

\title{Micro-transfer Printed Blue \mbox{InGaN} Lasers on Silicon Nitride Photonic Integrated Circuits}

\author[1,2]{Konstantinos~Akritidis}
\author[3]{Krzysztof~Gibasiewicz}
\author[1,2]{Han~Wang}
\author[3]{Iryna~Levchenko}
\author[1,2]{Max~Kiewiet}
\author[1,2]{Maximilien~Billet}
\author[3]{Mikołaj Chlipała}
\author[3]{Karolina~Peret-Malessa}
\author[2]{Pol~Van~Dorpe}
\author[3]{Henryk~Turski}
\author[1,2]{Bart~Kuyken}

\affil[1]{Photonics Research Group, INTEC Department, Ghent University -- imec, 9052 Ghent, Belgium}
\affil[2]{imec, Kapeldreef 75, 3001 Leuven, Belgium}
\affil[3]{Institute of High Pressure Physics Polish Academy of Sciences, PAS, 01-142 Warsaw, Poland}

\date{\normalsize Corresponding author: \texttt{Konstantinos.Akritidis@ugent.be}}

\begin{document}
\maketitle

\begin{abstract}
Expanding integrated photonics into the blue spectral range requires high-performance light sources, making the gallium nitride (GaN) material family indispensable. While silicon nitride (SiN) platforms offer a robust, CMOS compatible passive ecosystem for visible wavelengths, seamlessly integrating GaN lasers remains a major bottleneck. Conventional heterogeneous integration methods present distinct trade-offs: full-wafer bonding achieves high throughput but requires careful management of thermal and lattice mismatches across large areas, whereas flip-chip bonding ensures high yield through pretesting but is constrained by sequential processing speed. In this landscape, micro-transfer printing (MTP) emerges as a disruptive, material-efficient alternative, bypassing these limitations by combining high-density parallel integration with known-good-die selection. Applying MTP to GaN, however, presents a significant material challenge: due to its chemical inertness and strong III--N bonds, device release typically relies on electrochemical etching, which can compromise material quality. Here, we overcome this hurdle and demonstrate the first micro-transfer printed  blue lasers on a SiN platform. Using a heavily doped n-type sacrificial layer together with optimized electrochemical etching conditions, we release smooth-surfaced thin-film light sources from bulk GaN substrates. Following release, the devices are integrated and butt-coupled to SiN fork-shaped edge couplers, achieving high current densities exceeding 20 kA$/$cm$^2$ alongside lasing at 455~nm. These results  expand the visible integrated photonic toolkit and establish a framework for multi-wavelength integration, opening new avenues for next-generation technologies including flow cytometry, quantum computing, optical communications, and augmented/virtual reality.
\end{abstract}

% ================================================================
%  BODY
% ================================================================

\section{Introduction}

Visible-light integrated photonics marks a pivotal shift in modern optics, unlocking a new class of ultra-compact, high-efficiency, and low-noise optical systems \cite{tran_extending_2022, lu_emerging_2024, buzaverov_silicon_2024}. In quantum information science, the scaling of trapped-ion and neutral-atom systems heavily relies on precise visible wavelengths to transition from bulky, vibration-sensitive laboratory tables to robust, mass-manufacturable chip-scale architectures. For instance, in $^{88}$Sr$^+$ systems, ion loading and control require the integration of 405~nm and 461~nm lasers  \cite{niffenegger_integrated_2020}, while driving the 435.5~nm transition in ytterbium ions is crucial for developing optical atomic clocks \cite{stenger_absolute_2001}. Parallel to quantum advancements, the life sciences benefit immensely from visible photonic integration; specialized 488~nm light sources are vital for high-throughput biophotonics, enabling cellular-level biological sensing via lab-on-a-chip flow cytometry and fluorescence lifetime imaging \cite{kanno_high_throughput_2024, mckinnon_flow_2018, nedbal_timedomain_2015}. Furthermore, consumer and industrial markets demand efficient visible sources, where true 450~nm blue emitters are critically required to achieve the ultra-dense pixel architectures and high-brightness demands of next-generation augmented, virtual, and mixed reality (AR/VR/MR) displays \cite{shi_flat_panel_2025}. Concurrently, the 400-530~nm spectral window experiences minimal attenuation in water, positioning these compact visible sources as key enablers for high-bandwidth underwater optical wireless communications (OWC) and localized oceanic LiDAR systems \cite{wu_blue_2017, li_dual_wavelength_2020}. Realizing this vast spectrum of disruptive applications, however, hinges on the scalable and reliable integration of high-performance gallium nitride (GaN)-based light sources.

\begin{figure}[!b]
\centering\includegraphics[width=1.0\textwidth]{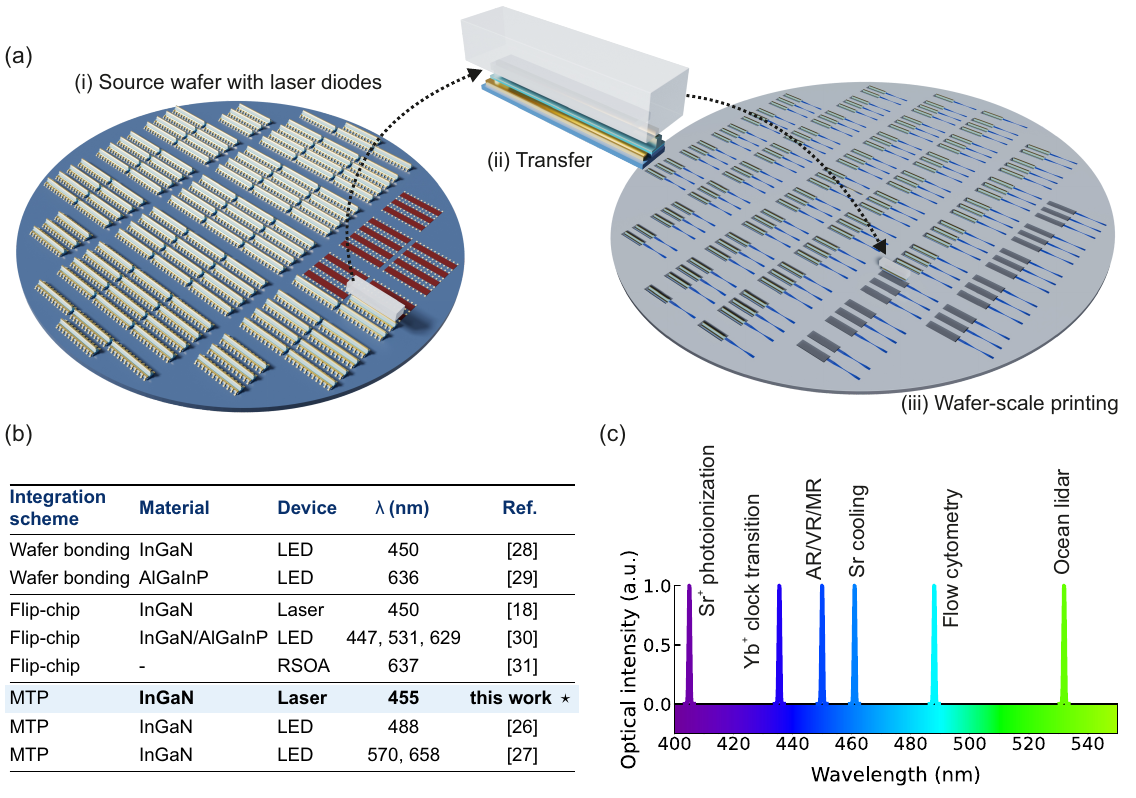}
\caption{(a)~Conceptual illustration of the micro-transfer printing technique. The source wafer contains thousands of pre-fabricated devices. An elastomer stamp picks up a laser bar (which can be parallelized into an array for high throughput) and transfers it to the silicon-based substrate. (b)~Representative examples of heterogeneously integrated light sources emitting in the visible spectral range (400-750 nm). (c)~Key technological domains unlocked by the scalable integration of GaN-based laser sources on PICs.}
\label{concept}
\end{figure}

The foundation of such on-chip systems rests on wide-bandgap dielectric passive waveguide platforms. While conventional silicon-on-insulator (SOI) technology dominates the telecommunication bands, silicon's narrow bandgap results in prohibitive optical absorption across the visible spectrum. Consequently, alternative dielectrics must be leveraged to provide low-loss routing frameworks. Among these, silicon nitride (SiN) has established itself as the premier candidate for photonic integrated circuits (PIC) owing to its ultra-wide bandgap, CMOS compatibility, seamless co-integration with electronics \cite{wan_integrating_2025}, and ultra-low propagation losses, reaching \mbox{sub-dB/cm} at 405 nm \cite{morin_cmos_foundry_based_2021}. Complementary foundry platforms utilizing materials such as aluminum oxide (AlOx) have also been developed, further extending operation into the deep blue and near-ultraviolet regime \cite{neutens_200_2025}.  However, because these wide-, indirect-bandgap dielectrics lack native optical gain, they remain intrinsically passive and incapable of light emission or amplification.

To bridge this material gap and introduce optical gain into passive dielectric platforms, several integration strategies have been explored. Monolithic growth on native GaN or sapphire substrates is commercially mature \cite{zhao_recent_2025}, but direct epitaxy on CMOS-compatible platforms like SiN or \mbox{AlOx}, despite the promising progress \cite{sun_room_temperature_2016}, still suffers from severe thermal/lattice mismatches and threading dislocation densities that significantly limit device yield. Hybrid die-to-die edge coupling offers superior thermal management and high optical output powers~\cite{corato_zanarella_widely_2023}, yet its low throughput hinders medium-to-high volume manufacturing. Consequently, heterogeneous integration has emerged as the leading paradigm to bridge passive and active functionalities. Flip-chip bonding enables the integration of pre-fabricated, pre-tested devices, as demonstrated with blue InGaN lasers on SiN interposer \cite{mu_hybrid_2026, wang_flip_hybrid_2026}, but its sequential nature (typically 100 laser units per hour \cite{marinins_wafer_scale_2022}) restricts its use to low-volume applications. Conversely, although wafer bonding achieves ultra-high throughput, it precludes pre-testing, suffers from substantial material waste, and remains in early development for blue GaN-on-SiN lasers~\cite{kamei_research_2020, liang_recent_2021, aggarwal_gan__si_2026}. Micro-transfer printing (MTP) offers a scalable alternative integration route by enabling highly parallel transfer of devices from their native substrates (Fig.~\ref{concept}(a)), while retaining compatibility with known-good-die pre-testing strategies \cite{roelkens_present_2024}. Furthermore, it maximizes material efficiency, a crucial factor for costly III-nitride platforms, and provides back-end-of-line compatibility, allowing device integration without introducing contamination risks to the ultra-clean fabrication lines.

While MTP has been successfully applied across various material platforms and device architectures \cite{yu_advancements_2025}, its extension to GaN-based optoelectronics remains exceptionally challenging. Unlike conventional III-V materials like gallium arsenide (GaAs) or indium phosphide (InP), that utilize mature, highly selective wet-chemical release layers, the extreme chemical inertness and strong covalent bonding of III--nitrides preclude standard isotropic wet underetching. This necessitates harsher release mechanisms, such as electrochemical underetching in acidic media. Consequently, MTP of GaN-based devices has so far been limited to light-emitting diodes (LEDs) \cite{chlipala_electrochemical_2025, lin_transfer_printed_2026}, which are significantly less demanding than laser systems, where optical facet quality, mirror coating preservation, and waveguide losses, are paramount to reaching lasing threshold.

In this work, we overcome this bottleneck by developing a robust release process tailored for GaN-based laser diodes, enabling the first proof-of-concept demonstration of micro-transfer printed blue lasers integrated on silicon nitride photonic integrated circuits via optimized fork-shaped edge couplers. These results establish an interdisciplinary milestone at the intersection of material engineering, parallel micro-assembly, and optoelectronics, paving the way for fully integrated visible photonic systems across quantum computing, augmented reality displays, and biosensing (Fig.~\ref{concept}(b, c)).

%%%%%%%%%%%%%%%%%%%%%%%%%%%%%%%%%%%%%%%%%%%%%%%%%%%%%%%%%%%%%%%%%%%%%%%%%%%%%%%%%%%
\nocite{gao_gan_2026}
\nocite{lee_wafer_bonded_2026}
\nocite{li_full_color_2022}
\nocite{mandal_silicon_2026}
\nocite{lin_transfer_printed_2026}

\section{Materials and methods}

\subsection{Electrochemical release of GaN-based devices}

Micro-transfer printing relies on growing functional devices on a sacrificial release layer that can be  selectively removed without damaging surrounding structures. Post-release, the devices remain suspended over their native substrate via dielectric tethers, designed to fracture deterministically at designated points during pick-up with a polymer stamp \cite{chen_micro_transfer_2026}. Extending this process to III-nitride devices, however, specifically for GaN-based laser diodes, presents a fundamental material challenge. Owing to the exceptionally strong Ga--N covalent bond, and the chemical inertness of III--nitrides, conventional wet chemical etchants cannot achieve the lateral etch required, necessitating the development of other methods such as electrochemical etching \cite{sawicka_air_cladding_2026, yao_monolithic_2026, harris_lee_porous_2025}.

\begin{figure}[!b]
\centering\includegraphics[width=1.0\textwidth]{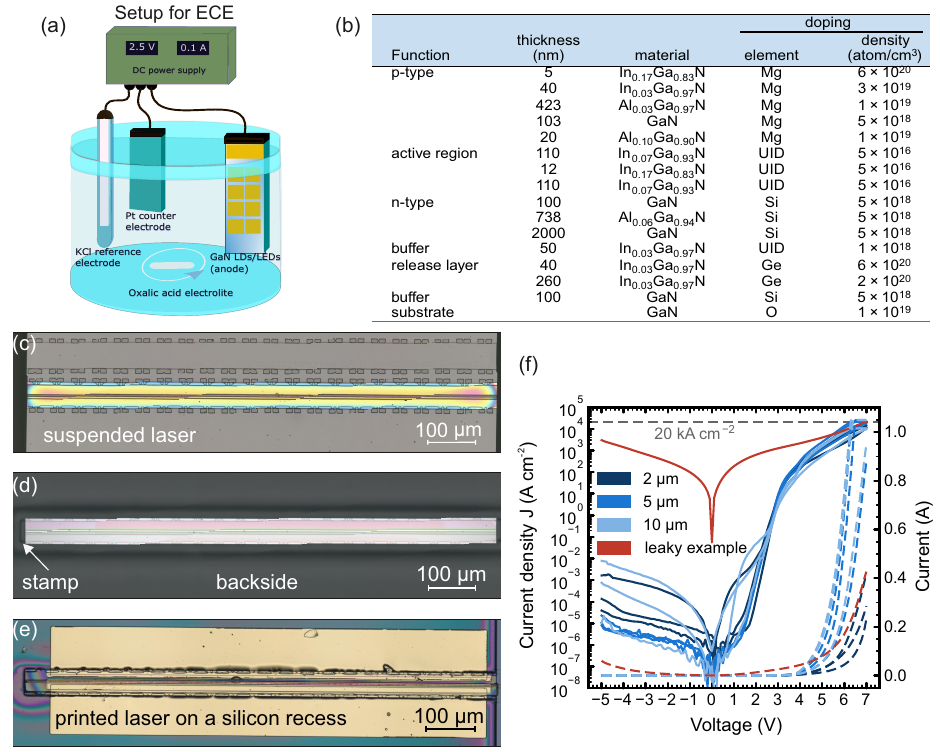}
\caption{(a)~Schematic of the setup used for the electrochemical underetching of the GaN-based active devices. (b)~Epitaxial layer stack of the GaN-based laser diodes, detailing layer composition and doping. (c-e)~Optical microscope images showing (c)~a suspended laser coupon with cleaved facets on the source wafer, (d)~the backside of the picked laser diode, (e)~the integrated laser. (f)~Current density and absolute current versus voltage characteristics of representative transferred lasers of different ridge widths.}
\label{underetching}
\end{figure}

The sacrificial layer removal procedure, which was reported in our previous work for the release and transfer of micro-LEDs \cite{chlipala_electrochemical_2025}, serves as the baseline for the laser diode process demonstrated here. Electrochemical underetching is performed in a three-electrode cell containing 0.3~M oxalic acid under constant magnetic agitation, as illustrated in Fig.~\ref{underetching}~(a). The GaN sample serves as the working anode, while a platinum mesh acts as the counter cathode. A reference electrode monitors the precise potential drop across the semiconductor-electrolyte interface to maintain controlled etching kinetics. Under an applied positive bias, holes are delivered driving oxidation and dissolution of the semiconductor according to \cite{harris_lee_porous_2025}:

$$\mathrm{2GaN +6h^+ \rightarrow 2Ga^{3+} + N_2 \uparrow.}$$

Electrochemical underetching is substantially more demanding for laser diodes than for LEDs, as it requires maintaining the integrity of several highly sensitive features, including the cavity facets, the long and narrow p-n junction mesa, and the processed ohmic contacts, while preventing electrolyte-induced damage throughout the release process. To achieve high lateral etch selectivity, while minimizing structural damage, a heavily germanium (Ge)-doped $\text{In}_\text{0.03}\text{Ga}_{\text{0.97}}\text{N}$ sacrificial release layer is implemented. Unlike conventional silicon doping, heavy Ge incorporation achieves ultra-high concentrations \cite{konczewicz_electrical_2022, chlipala_electrochemical_2025} without inducing severe lattice strain, significantly lowering the breakdown potential required.

The large disparity in free-carrier density between the InGaN:Ge release layer and the adjacent standard n--GaN cladding layers enables direct, highly selective electropolishing. The epitaxial stack along with the chemical composition and doping is detailed in Fig.~\ref{underetching}~(b). Notably, the release layer consists of two different doping levels with the highest density in the upper part of the layer to ensure the highest possible smoothness of the etched bottom surface, which is crucial for the subsequent integration step. 

Electrochemical etching was carried out on a 100 \textmu m-thick sample, mechanically thinned to enable reliable cleaving. The processed piece measured approximately 1~mm in width and 1-2~cm in length, and contained laser diodes with 1 mm cavity lengths defined by cleaved facets prior to etching. At an applied bias of 2.5 V, complete and residue-free underetching of the laser coupons was achieved within 5 hours, yielding an ultra-smooth release interface. Device encapsulation was realised using a tri-layer dielectric ($\text{SiO}_2/\text{SiN}/\text{SiO}_2$) deposited via plasma-enhanced chemical vapor deposition (PECVD). To direct the lateral etching front, a spin-coated photoresist layer was patterned to open a single longitudinal trench along one side of the laser bar, enforcing unidirectional electrolyte access. The fully released GaN laser with cleaved optical facets post photoresist stripping, is presented in Fig.~\ref{underetching}~(c). 

Due to the increased porosity associated with the lower-density dielectric deposition of our R\&D process environment, prolonged electrolyte exposure can result in localized chemical attack and delamination at the n-contact interface. This layer is substantially more vulnerable than the p-type metallization, which remains better protected atop the ridge waveguide. To circumvent this limitation, a post-integration metallization scheme was adopted: coupons were printed carrying solely the p-contact, deferring n-contact deposition to the post-transfer \mbox{Back-End-of-Line} processing stage. While this modular fabrication flow remains fully compatible with wafer-scale standard photonic integrated circuit manufacturing, optimizing the dielectric film density in future iterations will simplify post-processing, potentially reduce n-type contact resistance, and crucially, will enable pre-testing and known-good-die screening prior to integration to further maximize yield.

Inspection of the picked coupon's backside, shown in Fig.~\ref{underetching}~(d), confirms a very smooth, residue-free surface. Minimizing backside roughness is critical, as it maximizes interfacial contact area and Van der Waals adhesion with the target substrate during printing. This high surface quality enables direct bonding, while in the case of adhesive-assisted printing, it significantly reduces the required adhesive layer thickness, thereby mitigating thermal constraints caused by the inferior thermal conductivity of polymer adhesives such as DVS-BCB (divinylsiloxane-bis-benzocyclobutene).

To evaluate the electrical quality of the transfer-printed devices, DC current-voltage characterization was performed on the integrated GaN-on-silicon laser diodes (Fig.~\ref{underetching}~(e)). The coupons featured ridge widths of 2, 5, and 10 \textmu m across a cavity length of approximately 1~mm. The resulting characteristics, plotted as both absolute current and current density versus voltage, are presented in Fig.~\ref{underetching}~(f). The devices exhibit high-current operation with current densities exceeding 20~$\text{kA/cm}^2$ at a forward bias of about 7~V, limited only by the maximum output (1050~mA) of the source meter. The  reverse-bias leakage observed in the non-ideal device is primarily attributed to metal overlap along the ridge perimeter caused by a challenging lithographic lift-off process, which can create localized parasitic current paths across the p-n~junction. 
 
\subsection{Design and optimization of SiN fork-based edge couplers}

\begin{figure}[!t]
\centering\includegraphics[width=1.0\textwidth]{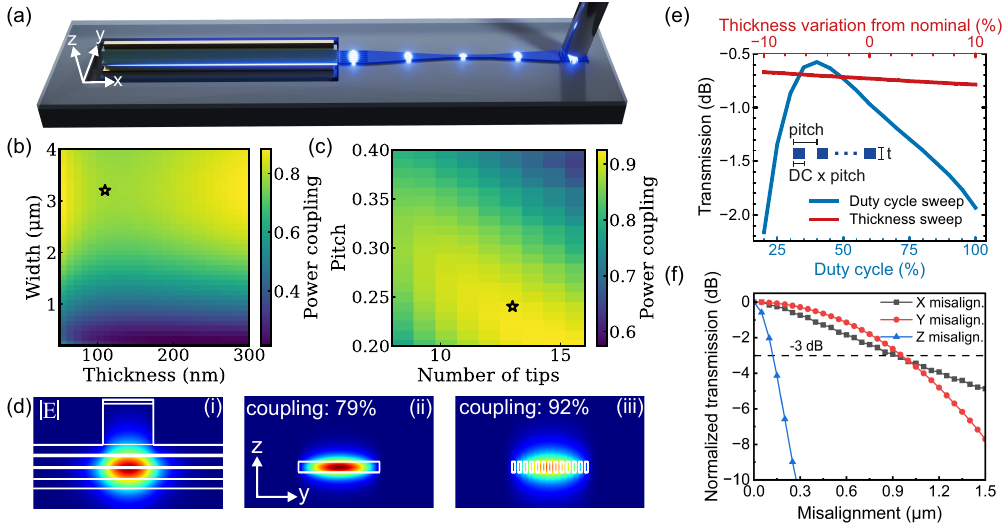}
\caption{(a)~Schematic illustration of the heterogeneously integrated, butt-coupled blue laser on silicon nitride. (b, c)~2D finite-difference eigenmode colormaps showing the power coupling at a wavelength of 455~nm between a 2~\textmu m wide ridge laser diode and: (b)~a SiN waveguide for different cross sections; (c)~a SiN fork-shaped structure with varying pitches and number of tips at a 50\% duty cycle. For a SiN thickness of 110 nm, the power coupling in the fork-shaped structure increases significantly (values shown by the star symbols). (d)~Mode profiles of the individual structures, with the calculated power coupling values annotated inside the plots. (e)~Simulated 3D~FDTD transmission of the GaN-to-SiN coupling as a function of duty cycle and thickness variation from the nominal value. (f)~Normalized 3D FDTD transmission  of the GaN-to-SiN coupling as a function of misalignment along the x-, y-, and z-axes during the integration process.}
\label{simulations}
\end{figure}

Photonic integration of the GaN blue laser with the SiN photonic integrated circuit is realized via a butt-coupling architecture. The laser coupon is integrated on a silicon recess, allowing precise vertical alignment between the active region and the SiN waveguide core, as schematically illustrated in Fig.~\ref{simulations}~(a). While evanescent coupling is widely employed at near-infrared wavelengths, its implementation at visible frequencies is constrained by the refractive index mismatch between the SiN (n $\approx$ 2) and the GaN (n $\approx$ 2.7), which precludes satisfying the phase-matching conditions directly. Although intermediate dielectric layers have been used at 980~nm to bridge this index gap between GaAs and SiN \cite{akritidis_heterogeneous_2026}, identifying suitable materials at 450~nm is exceptionally challenging due to strict refractive index constraints and optical absorption. Alternatively, while adiabatic coupling can be achieved without an interlayer by aggressively thinning the laser's n-contact cladding and defining ultra-sharp tapers, this approach increases n-contact sheet resistance, thereby degrading electrical performance, and introduces sensitive fabrication steps that directly counteract the high-yield, wafer-scale advantages of micro-transfer printing. In contrast, butt-coupling circumvents these material limitations through a universal interface that simultaneously leverages direct contact with the silicon substrate for superior heat dissipation, bypassing the high thermal resistance of the underlying buried oxide layer \cite{kiewiet_microtransfer_2026, uzun_integration_2023}.

To optimize optical power transfer across the interface, the power coupling between the GaN laser output and SiN strip waveguides across a broad parameter space was evaluated, using finite-difference eigenmode (FDE) mode overlap simulations. These simulations assume perfect alignment and zero coupling gap, and account only for mode and index mismatch, thus providing an upper bound on the achievable performance. Fig.~\ref{simulations}~(b) maps the coupling efficiency as a function of waveguide core thickness (50-300~nm) and width (0.1-4~\textmu m). High coupling values (>85\%) are restricted to two extreme regimes: Ultra-thin (50-60~nm) or thick (280~nm) cores, leaving intermediate film thicknesses considerably mismatched. Although very thin films expand the mode vertically yielding high coupling efficiency, a strategy previously demonstrated in GaN-based flip-chip integrated lasers \cite{mu_hybrid_2026}, they suffer from poor mode confinement and require excessively large bend radii, compromising circuit footprint. Conversely, thick SiN layers at 455~nm operation introduce higher-order mode excitation and polarization instability, rendering single-mode operation difficult to sustain.

To resolve this performance trade-off, a multi-tip fork-shaped SiN edge coupler \cite{kluge_flip_chip_integrated_2022, wang_compact_2025} is implemented that expands the guided optical mode at moderate core thicknesses. Fixing the film thickness within the intermediate regime (110 nm) at a 50\% duty cycle, we systematically varied the tip pitch and tip count, achieving a power coupling of nearly 92\% (Fig.~\ref{simulations}~(c)). To maintain manufacturing compatibility with standard deep ultraviolet (DUV) lithography processes, a minimum pitch of 240 nm was enforced within the parameter space, ensuring a minimum feature size of 120 nm, yielding in turn an optimal number of thirteen tips. Modal profiles for the GaN laser emission, the standard 110 nm SiN waveguide, and the engineered fork coupler are compared in Fig.~\ref{simulations}~(d)(i-iii). By locally tailoring the effective index, the fork coupler effectively delocalizes the optical field to match the laser profile, delivering high coupling efficiency without incurring footprint penalties or higher order mode excitation. Furthermore, this multi-tip configuration provides a strong starting geometry that can be further optimized via inverse-design algorithms to enhance coupling efficiency and manufacturing tolerances.

To evaluate the robustness of the coupler against structural and positional, integration-induced, variations, 3D finite-difference time-domain (FDTD) sensitivity simulations were conducted. Specifically, we investigated variations in the geometry of the fork-shaped coupler, namely its thickness and duty cycle, as well as positional alignment during integration. The corresponding results are depicted in Figs.~\ref{simulations}~(e) and (f). In all cases, a wavelength of 455 nm is considered and a BCB planarization is assumed to have taken place to fill any gaps between the integrated laser and the SiN waveguide. The layer thickness was varied by $\pm 10\%$, conservatively accounting for the typical $\approx5\%$ thickness tolerance provided by mature foundries for SiN deposition on 200~mm wafers \cite{akritidis_heterogeneous_2026}. Over this range, the transmission from the laser diode to the fundamental transverse electric (TE) mode of the coupler remains highly robust. Next, the duty cycle of the fork's tips was varied from 20\% to 100\% (where 100\% corresponds to a standard straight waveguide). Such duty-cycle variations in fabrication typically originate from lithographic linewidth fluctuations, optical proximity effects, or over/under-etching. As expected, increasing the duty cycle reduces mode expansion, leading to poorer coupling efficiency. Conversely, reducing the duty cycle below 50\% initially yields a slight improvement in transmission, followed by a rapid drop-off at lower values caused by an increased vertical mode expansion alongside a decreased lateral mode size, degrading the spatial mode overlap with the laser profile. This behavior introduces an asymmetric tolerance profile, with a maximum additional loss of 1.5~dB over the tested range. Importantly, while this parameter sweep spans a wide duty-cycle range to offer physical insights, realistic fabrication variation is expected to remain within 20~nm, maintaining a duty cycle close to the 40-60\%. Overall, these results demonstrate that the proposed fork-shaped coupler achieves a nominal simulated transmission of roughly -0.7~dB (85\%) while remaining resilient to standard fabrication tolerances, ensuring uniform wafer-scale performance.

For the assessment of the integration robustness, we investigated positional misalignments during the micro-transfer printing process up to a maximum displacement of 1.5 \textmu m. While a 1.5 \textmu m alignment window is representative of sample-scale printing with an array of transferred devices, state-of-the-art wafer scale printers achieve a 3$\sigma$ accuracy within 500 nm across full 200~mm wafers \cite{zheng_micro_transfer_2026}. For this study, the transmission efficiency was normalized against the ideal, perfectly aligned case (which corresponds to a transmission of roughly 85\%) to isolate the excess loss introduced during integration. For in-plane (x- and y-axis) misalignments, a displacement of 500~nm introduces an excess loss of approximately 1.5~dB, which escalates beyond 4~dB as the offset approaches 1.5 \textmu m. Conversely, the coupler exhibits significantly higher sensitivity to vertical (z-axis) misalignments, which can stem from epitaxial layer thickness variations, recess etch depth non-uniformity, BCB bonding layer variations, or buried/top oxide thickness fluctuations. A vertical offset of merely 150 nm incurs a 3~dB penalty, with losses reaching 10~dB near a 300~nm displacement. This high vertical sensitivity is consistent with previous demonstrations of flip-chip integrated blue lasers on SiN-based inverse tapers, where a 3~dB loss at a 200 nm vertical misalignment was reported~\cite{mu_hybrid_2026}.

This pronounced vertical sensitivity originates from two main factors. First, operating at a short wavelength ($\lambda$ = 455 nm) results in a tightly confined optical mode profile that is naturally more sensitive to offsets. Second, the fork-shaped geometry was specifically optimized to maximize spatial mode overlap. This introduces a trade-off: designs optimized for maximum coupling efficiency inherently experience larger relative loss penalties under misalignment compared to design variations with broader, relaxed mode profiles that sacrifice peak efficiency for displacement tolerance. While in-plane tolerances are well within the capabilities of modern MTP tools, mitigating vertical variation, either through tighter process control or by exploring vertically tolerant coupler topologies, remains crucial for high-yield wafer-scale integration.

\subsection{Target preparation and laser integration}

\begin{figure}[!t]
\centering\includegraphics[width=0.5\textwidth]{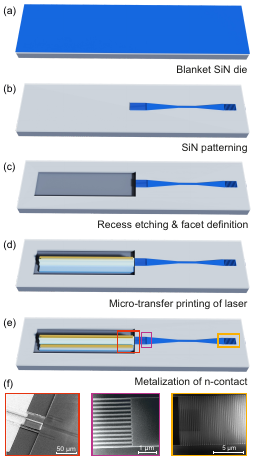}
\caption{Fabrication of the micro-transfer printed blue lasers. (a)~Blanket SiN layer overlying thermal oxide on a Si substrate. (b)~Patterning of the SiN photonic circuitry, including the fork-shaped waveguide and grating coupler, and deposition of high-density oxide top cladding. (c)~Deep recess etching reaching $\approx1$ \textmu m deep into the silicon substrate. During this step, the SiN facets are defined. (d)~Spin coating of adhesive BCB and integration of the GaN-based blue laser diode. (e)~BCB curing followed by metallization of the n-contact on the transferred laser coupon. In this schematic the top oxide cladding and the BCB layer are omitted for clarity. (f) Scanning electron microscope images displaying (from left to right) the integrated butt-coupled laser, and representative test structures for the fork-shaped SiN edge coupler and the output grating coupler.}
\label{fabrication}
\end{figure}

The simplified fabrication process flow for integrating blue GaN-based lasers onto the SiN platform via MTP is schematically illustrated in Fig.~\ref{fabrication}. Fabrication begins on a silicon substrate with 2~\textmu m of thermal oxide and a 110~nm low pressure chemical vapor deposition SiN layer~(a). The photonic circuit, comprising the fork-shaped coupler, passive routing waveguides, and out-coupling grating coupler, is patterned using electron beam lithography and dry etching, followed by the deposition of an inductively coupled plasma chemical vapor deposition top oxide cladding~(b). To prepare the target for laser integration, a deep dry etch is executed through the dielectric stack and into the silicon substrate using a chromium hard mask. This step simultaneously creates the precise recess for micro-transfer printing, monitored through profilometry and reflectometry, and defines the SiN coupling facets~(c). Post the hard mask removal, a thin BCB adhesive layer is spin-coated,  pre-fabricated InGaN laser coupons are transfer-printed into the target recess, and BCB curing is performed~(d). Finally, the n-type metallization is deposited through sputtering (Ti/Pt/Au), followed by a thick top-metallization step to define the contact pads for electrical characterization~(e).

Fig.~\ref{fabrication}~(f) shows scanning electron microscope (SEM) images of the butt-coupled laser integrated via micro-transfer printing, together with the SiN fork-shaped edge coupler and output grating coupler. The images confirm high-precision in-plane alignment of the printed laser, but also reveal that the edge coupler duty cycle deviates from the 50~\% design target. Electron-beam lithography proximity effects together with over-etching  caused a shift from the target 50\% duty cycle to a measured 35\%, which contributes to a negligible increase in coupling efficiency. At the coupling interface, a 200~nm BCB layer remains from the prior adhesive spin-coating process. This thin layer introduces minimal absorption or geometric mismatch penalty while simultaneously acting as an intermediate index-matching layer ($n\approx1.56$) that suppresses reflection compared to an air gap. Crucially, focused ion beam cross-sectional analysis confirmed that the significant BCB pooling at the facets reported in prior O-band  implementations \cite{uzun_integration_2023} was not present in these samples, leaving only the uniform thin layer. To achieve completely material-free optical facets in future iterations, photosensitive BCB could be adopted to selectively clear the coupling interface \cite{kiewiet_microtransfer_2026}. Furthermore, functional contact metallization was successfully achieved across the high-topography coupons without a dedicated planarization step, confirming robust step coverage across the coupons' edges and validating a simplified fabrication process flow for die-level prototyping.

\section{Results}

\begin{figure}[!t]
\centering\includegraphics[width=1.0\textwidth]{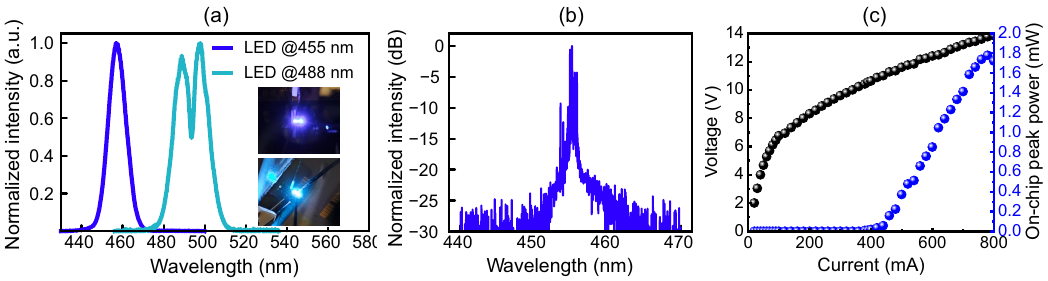}
\caption{Electro-optical characterization of the integrated InGaN laser diodes. (a)~Normalized electroluminescence spectra of two laser diodes operating in the spontaneous emission (LED) regime (95\% duty cycle, I~=~95~mA, OSA resolution of 2 nm), showing waveguide-coupled emission near 455~nm and 488~nm. Inset: Photographs of the operating devices. (b)~Normalized emission spectrum of the laser diode emitting at 455 nm in pulsed mode (0.1\% duty cycle, I~=~740~mA, OSA resolution of 100 pm). (c)~Light-current-voltage characteristics of the butt-coupled laser.}
\label{characterization}
\end{figure}

Following integration, electro-optical characterization was performed by electrically contacting the printed lasers with probes, coupling the emitted light into the SiN waveguides, collecting the output via a single-mode fiber aligned to the output grating couplers, and directing it to a power meter and an optical spectrum analyzer (OSA). Under continuous-wave (CW) or high-duty-cycle excitation, pronounced thermal self-heating combined with high threshold current density prevented the laser diodes from reaching lasing, resulting in spontaneous emission (LED-mode regime). To validate the versatility of this platform, devices fabricated from material with different indium concentration in the InGaN quantum wells, were integrated and characterized. As shown in Fig.~\ref{characterization}(a), waveguide-coupled spontaneous emission was successfully demonstrated near both 455 nm and 488 nm, measured at 2~nm OSA resolution. The non-gaussian  emission profile at 488 nm stems from the spectral envelope of the grating coupler, which was nominally designed for 455 nm and required angular tuning to capture the longer wavelength. Notably, when suppressing thermal loading by operating in low-duty-cycle pulsed mode (0.1 \% duty cycle, 900 \textmu s period), lasing was achieved at 455 nm. High-resolution spectral measurements (100~pm OSA resolution) and light-current-voltage (L--I--V) performance are shown in Figs.~\ref{characterization}(b) and (c), respectively. The device exhibits a threshold current of 450~mA and an on-chip peak power of approximately 1.8 mW within the SiN waveguide, which corresponds to an average measured on-chip power of 1.8 \textmu W at 0.1 \% duty cycle, de-embedded from the 13.5 dB grating coupler insertion loss.

\begin{figure}[!b]
\centering\includegraphics[width=0.9\textwidth]{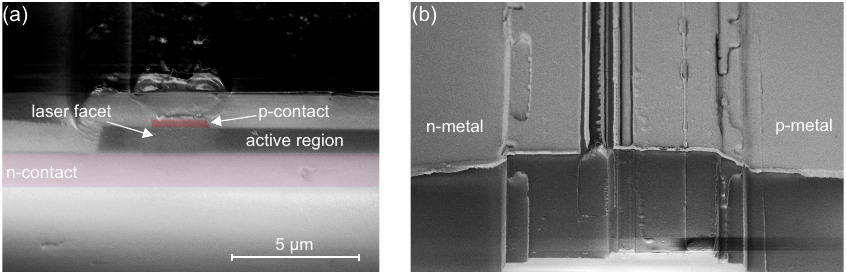}
\caption{(a)~SEM (false-colored) cross-sectional image of the integrated laser's back facet. (b)~Zoomed out SEM of the structure.}
\label{SEM_facet}
\end{figure}
The transition from spontaneous emission under DC operation to clear lasing under low-duty-cycle pulsed excitation provides an important proof-of-concept demonstration of the first transfer-printed GaN-based laser integrated onto a SiN photonic platform. Although pulsed lasing was achieved, continuous wave (CW) emission was not observed: the relatively high pulsed threshold current (450 mA), driven by optical losses in the cavity, together with the high series resistance, results in significant power dissipation, making thermal management the primary roadblock to CW operation. Under CW drive, this dissipation raises the junction temperature, which, as is well documented for GaN-based laser diodes \cite{pourhashemi_pulsed_2013, ryu_investigation_2017}, further increases the required threshold current through reduced differential efficiency, carrier leakage and other temperature-dependent loss mechanisms. In particular, the uncoated laser facets could be partially affected by the underetching process used for coupon release, as suggested by the defects on the facet (Fig.~\ref{SEM_facet} (a)), introducing additional mirror losses that could elevate the threshold current. Future implementations could address this by depositing facet coatings prior to the release process, which would both protect the facets during underetching and increase their reflectivity, thereby reducing cavity losses. Ultimately, the combination of these losses and self-heating under CW operation likely prevented the device from achieving the lasing condition, despite successful electrical injection. Importantly, these limitations are not intrinsic to the micro-transfer printing platform itself: as shown in Fig.~\ref{SEM_facet} (b), the printed device remains mechanically intact, with any degradation appearing localized to the facet surface. The demonstration of robust pulsed lasing, coupling into the SiN waveguide, and milliwatt-level peak on-chip power establishes a strong foundation for future optimization of thermal management, facet quality, and cavity design, paving the way toward low-threshold and continuous wavelength operation in subsequent generations of transfer-printed GaN on photonic integrated circuits.

\section{Conclusion}

The scalable integration of III-nitride sources marks a crucial step toward compact, visible-light photonic integrated circuits. Through electrochemical underetching, GaN-based laser diodes are successfully released and micro-transfer printed onto a SiN platform, achieving high current densities exceeding 20 kA$/$cm$^2$. By implementing fork-shaped SiN edge couplers, simulations demonstrate a transmission exceeding 85\% at 455 nm, confirming that sufficient mode expansion can be achieved without relying on ultra-thin waveguides. Building on these advances, waveguide-coupled spontaneous emission from two epitaxial structures (emitting at 455 nm and 488 nm), alongside pulsed-mode lasing at 455 nm, is shown, establishing the first demonstration of micro-transfer printed blue lasers on silicon nitride, an achievement at the intersection of material engineering,  heterogeneous assembly, and photonics.

To fully realize the potential of this approach, future efforts will focus on the optimization of the fabrication process to transition from proof-of-concept demonstrations to  high-yield, scalable manufacturing. By applying protective, high reflectivity mirror coatings prior to the release process, together with the implementation of heat-spreading structures for enhanced thermal management, robust continuous-wave operation with higher output powers can be unlocked. Paired with tailoring the silicon nitride cavity toward single-mode or tunable lasing, these advances will deliver the spectral control and performance required for next-generation visible-light applications, ranging from AR/VR displays and quantum technologies to optical communications.

\section*{Acknowledgments}
We acknowledge funding from the Horizon Europe program of the European Union (Grant Agreement ID: 101070622). 

\section*{Data availability statement}
The data that support the findings of this study are available from the corresponding author upon reasonable request.

\section*{Disclosures}
The authors declare no conflicts of interest.

\bibliography{references}

\end{document}